\documentclass [prl,aps,letterpaper,onecolumn,superscriptaddress,amsmath,amssymb,floatfix,longbibliography] {revtex4-2}

\usepackage{physics}
\usepackage{graphicx}

\usepackage{hyperref}

\renewcommand{\thefootnote}{\fnsymbol{footnote}}

\newcommand{\siqse}{Shenzhen Insititute for Quantum Science and Engineering, Southern University of Science and Technology, Shenzhen 518055, China}
\newcommand{\physustech}{Department of Physics, Southern University of Science and Technology, Shenzhen 518055, China}
\newcommand{\gdpkl}{Guangdong Provincial Key Laboratory of Quantum Science and Engineering, Southern University of Science and Technology, Shenzhen 518055, China}
\newcommand{\szkl}{Shenzhen Key Laboratory of Quantum Science and Engineering, Southern University of Science and Technology, Shenzhen 518055, China}

\begin{document} 
\title{Suppressing Coherent Two-Qubit Errors via Dynamical Decoupling}

\author{Jiawei Qiu}
\thanks{These authors contributed equally.}
\affiliation{\siqse}\affiliation{\physustech}
\author{Yuxuan Zhou}
\thanks{These authors contributed equally.}
\affiliation{\siqse}\affiliation{\physustech}

\author{Chang-Kang Hu}
\affiliation{\siqse}\affiliation{\gdpkl}\affiliation{\szkl}
\author{Jiahao Yuan}
\affiliation{\siqse}\affiliation{\physustech}
\author{Libo Zhang}
\affiliation{\siqse}\affiliation{\gdpkl}\affiliation{\szkl}

\author{Ji Chu}
\affiliation{\siqse}
\author{Wenhui Huang}
\affiliation{\siqse}\affiliation{\physustech}
\author{Weiyang Liu}
\affiliation{\siqse}\affiliation{\gdpkl}\affiliation{\szkl}
\author{Kai Luo}
\affiliation{\siqse}\affiliation{\physustech}
\author{Zhongchu Ni}
\affiliation{\siqse}\affiliation{\physustech}
\author{Xianchuang Pan}
\affiliation{\siqse}
\author{Zhixuan Yang}
\affiliation{\siqse}
\author{Yimeng Zhang}
\affiliation{\siqse}

\author{Yuanzhen Chen}
\affiliation{\siqse}\affiliation{\physustech}\affiliation{\gdpkl}\affiliation{\szkl}
\author{Xiu-Hao Deng}
\affiliation{\siqse}\affiliation{\gdpkl}\affiliation{\szkl}
\author{Ling Hu}
\affiliation{\siqse}\affiliation{\gdpkl}\affiliation{\szkl}
\author{Jian Li}
\affiliation{\siqse}\affiliation{\gdpkl}\affiliation{\szkl}
\author{Jingjing Niu}
\affiliation{\siqse}\affiliation{\gdpkl}\affiliation{\szkl}
\author{Yuan Xu}
\affiliation{\siqse}\affiliation{\gdpkl}\affiliation{\szkl}
\author{Tongxing Yan}
\affiliation{\siqse}\affiliation{\gdpkl}\affiliation{\szkl}
\author{Youpeng Zhong}
\affiliation{\siqse}\affiliation{\gdpkl}\affiliation{\szkl}
\author{Song Liu}
\thanks{lius3@sustech.edu.cn}
\affiliation{\siqse}\affiliation{\gdpkl}\affiliation{\szkl}
\author{Fei Yan}
\thanks{yanf7@sustech.edu.cn}
\affiliation{\siqse}\affiliation{\gdpkl}\affiliation{\szkl}
\author{Dapeng Yu}
\affiliation{\siqse}\affiliation{\physustech}\affiliation{\gdpkl}\affiliation{\szkl}

\begin{abstract} 
\textbf{Scalable quantum information processing requires the ability to tune multi-qubit interactions. This makes the precise manipulation of quantum states particularly difficult for multi-qubit interactions because tunability unavoidably introduces sensitivity to fluctuations in the tuned parameters, leading to erroneous multi-qubit gate operations. The performance of quantum algorithms may be severely compromised by coherent multi-qubit errors. It is therefore imperative to understand how these fluctuations affect multi-qubit interactions and, more importantly, to mitigate their influence. In this study, we demonstrate how to implement dynamical-decoupling techniques to suppress the two-qubit analogue of the dephasing on a superconducting quantum device featuring a compact tunable coupler, a trending technology that enables the fast manipulation of qubit--qubit interactions. The pure-dephasing time shows an up to $\sim$14 times enhancement on average when using robust sequences. The results are in good agreement with the noise generated from room-temperature circuits. Our study further reveals the decohering processes associated with tunable couplers and establishes a framework to develop gates and sequences robust against two-qubit errors.}
\end{abstract}

\maketitle
\renewcommand{\thefootnote}{\arabic{footnote}}

High-fidelity quantum operations are key to scalable quantum information processing. For example, in the gate-based model, two-qubit gates currently constitute the performance bottleneck for noisy intermediate-scale quantum devices~\cite{preskill_quantum_2018} as a result of nonidealities in physical devices, which make it extremely difficult to precisely manipulate qubit--qubit interactions. The idea of using a tunable coupler to enable independent control over the coupling strength has proven effective in recent experiments on superconducting qubit platforms~\cite{arute_quantum_2019,foxen_demonstrating_2020,kandala_demonstration_2020,li_tunable_2020,collodo_implementation_2020,xu_highfidelity_2020,sung_realization_2020,stehlik_tunable_2021, xu_realisation_2021}, resulting in substantial improvements in the gate performance.

Despite successful demonstrations, the cost of using a tunable coupler, e.g., the new decohering processes it introduces, requires further investigation and reducing this cost remains a challenge. In fact, fast modulation of the coupling unavoidably leads to strong sensitivity to the biasing parameter of the coupler, which, in the case of superconducting qubits, is often the magnetic flux threading a superconducting quantum interference device (SQUID) loop. Therefore, any flux fluctuations, e.g., from instrumental instabilities~\cite{fried_assessing_2019}, electromagnetic interference, and the ubiquitous $1/f$ flux noise in solid-state devices~\cite{wellstood_lowfrequency_1987,yoshihara_decoherence_2006,bialczak_1f_2007}, will lead to diffusion in the qubit--qubit interaction rates, compromising the two-qubit gate performance. Unfortunately, such coherent or unitary errors are often overlooked in conventional benchmarking tests based on random quantum circuits, but may build up much faster in circuits with repeated operations~\cite{kjaergaard_programming_2020,arute_hartree-fock_2020,karamlou_analyzing_2020}, as required by most quantum algorithms. It is therefore insufficient to consider only the digitized errors. More importantly, even though studies have been done on predicting surface code~\cite{bravyi_quantum_1998,fowler_surface_2012} performance with single-qubit coherent errors~\cite{bravyi_correcting_2018}, two-qubit Pauli errors~\cite{fowler_surface_2012}, and two-qubit crosstalk errors (static)~\cite{huang_alibaba_2020} respectively, it is still unclear how error-correction codes will be affected by coherent two-qubit errors. These errors post additional challenges to developing fault-tolerant quantum computers.

It is the need to suppress these coherent two-qubit errors that motivates studies of the noise model and related decohering processes, as well as the search for open-loop quantum control strategies, such as dynamical-decoupling (DD) techniques for error mitigation\cite{viola_dynamical_1998,suter_colloquium_2016}. These techniques have been successfully implemented in various quantum systems~\cite{du_preserving_2009,ryan_robust_2010,biercuk_optimized_2009,bluhm_dephasing_2011,piltz_protecting_2013a,zhang_experimental_2015}. In superconducting quantum circuits, decoherence mitigation has been demonstrated in a single-qubit system~\cite{bylander_noise_2011,guo_dephasinginsensitive_2018,pokharel_demonstration_2018,sung_nongaussian_2019,souza_process_2020,jurcevic_demonstration_2021,chen_exponential_2021} and in a system consisting of a qubit and a spurious two-level system~\cite{gustavsson_dynamical_2012}. The ``net-zero'' technique~\cite{rol_fast_2019} which takes advantage of specific system properties to cancel noise when performing two-qubit gate is reminiscent of dynamical correction. However, general implementation of DD with two-qubit interactions has yet to be shown. 

In this study, we extend dynamical-decoupling techniques to the realm of two-qubit interactions featuring a tunable coupler that can continuously tune the excitation-swapping rate (absolute value) between two qubits from exactly zero to $\sim$100~MHz. In the two-level subspace spanned by the Bell states, the tunable coupler introduces a strong dephasing effect to the swapping evolution (which we refer to as swap-dephasing) as a result of low-frequency flux noise. By establishing the framework for full control over this subspace qubit, we implemented analogues of various DD sequences, demonstrating an elongation of the pure-dephasing time up to approximately 14 times. An error analysis shows that the relevant flux noise originates from room-temperature electronics and ground loops.

\begin{figure}[bth]
\centering
\includegraphics[scale=0.8]{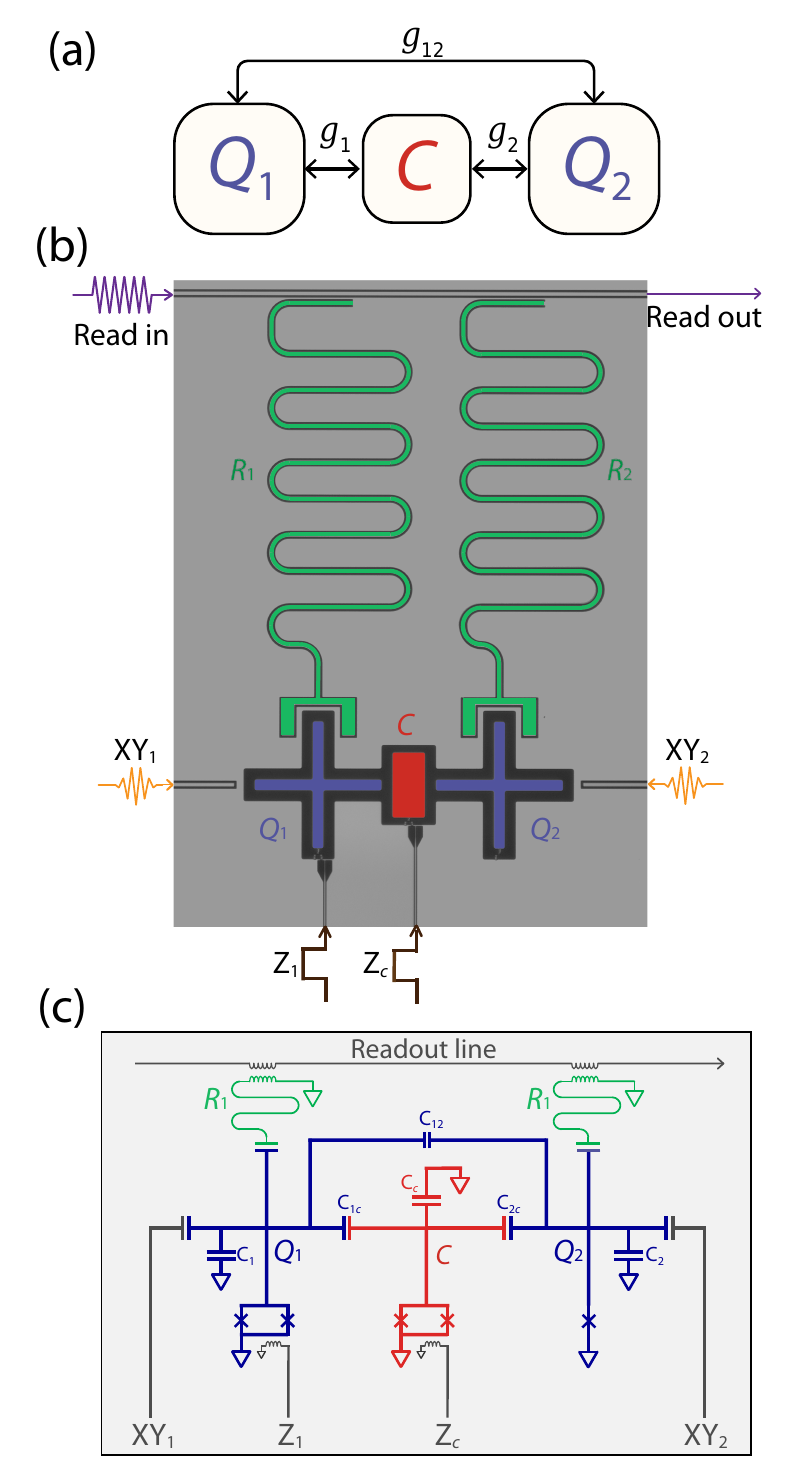}
\caption{Device design. 
(a) Schematic of the architecture with a tunable coupler between two qubits and (b) micrograph of the physical device implemented with superconducting quantum circuits. Two Xmon qubits (blue) and a tunable coupler (red) are capacitively connected via their shunt capacitor pads. Each qubit is connected to a quarter-wavelength transmission-line resonator (green) for dispersive readouts. The readout signal is transmitted through a common transmission line coupled to the two resonators. Both qubits have local $XY$ drive lines for performing single-qubit operations. The left qubit is tunable and has an additional flux or $Z$ bias line. The coupler also has its own $Z$ bias line. 
(c) Circuit diagram of the device. The qubit--coupler coupling capacitance $C_\mathrm{1C} = C_\mathrm{2C} = 3.3$~fF, and the qubit--qubit direct-coupling capacitance $C_\mathrm{12} = 0.16$~fF.}
\label{fig1}
\end{figure}

The device used in our experiment was made of aluminium on sapphire, with the metal-layer layout shown in Fig.~\ref{fig1}(b). Our tunable coupler, sandwiched by two Xmon qubits~\cite{barends_coherent_2013}, is a split transmon qubit that is capacitively shunted by a square-shaped pad to the ground. Such a compact design is advantageous for scaling up while allowing both strong nearest-neighbour qubit--coupler coupling ($g_\mathrm{1c}/2\pi = 122$~MHz, $g_\mathrm{2c}/2\pi = 105$~MHz) and strong next-nearest-neighbour direct qubit--qubit coupling ($g_{12}/2\pi = 12$~MHz), which are crucial for achieving high-fidelity two-qubit gates, as previously demonstrated with this same device~\cite{xu_highfidelity_2020}. The maximum frequency of the coupler $\omega_\mathrm{c}(\Phi_\mathrm{c}\!=\!0)/2\pi \approx 8.8$~GHz, where $\Phi_\mathrm{c}$ is the magnetic flux threading the SQUID loop (Fig.~\ref{fig1}(b) inset). One qubit ($Q_1$) is tunable, with its maximum frequency at $\omega_1(\Phi_1\!=\!0)/2\pi=5.27$~GHz ($\Phi_\mathrm{1}$ indicates the loop flux of the qubit), and the other qubit has a fixed frequency at $\omega_2/2\pi=4.62$~GHz. Both qubits have local $XY$ drive lines (bandwidth: 7.5~GHz), and both $Q_1$ and the coupler have local $Z$ bias lines (bandwidth: 500~MHz) for frequency tuning. Each qubit is connected to a quarter-wavelength resonator at about 7~GHz for dispersive readout. The chip is packaged and mounted in a dilution refrigerator (base temperature: $\sim$10~mK) during measurements. More details concerning the device design and the measurement setup can be found in Fig.~\ref{fig1}(c) and the supplementary material~\cite{supplement}.

Our design enables the tunable-coupling scheme previously proposed in Ref.~\cite{yan_tunable_2018}, in which the net qubit--qubit coupling strength $g$ can be tuned continuously from positive to negative by controlling the coupler frequency $\omega_\mathrm{c}$ via the coupler flux $\Phi_\mathrm{c}$. The truncated Hamiltonian in the subspace formed by $\ket{01}$ and $\ket{10}$ can be expressed as
\begin{equation}
    \mathcal{H} = \frac{1}{2} \Delta(\Phi_1)\big(\ket{10}\bra{10} - \ket{01}\bra{01}\big) + g(\Phi_\mathrm{c}) \big(\ket{10}\bra{01} + \ket{01}\bra{10}\big),
    \label{hamiltonian_1}
\end{equation}
where $\Delta(\Phi_1)=\omega_1(\Phi_1)-\omega_2$ is the qubit--qubit detuning. Equation \ref{hamiltonian_1} suggests that qubit excitation will cycle between $\ket{10}$ and $\ket{01}$ at the frequency $\Omega(\Phi_\mathrm{c})=2g(\Phi_\mathrm{c})$ when $\Delta=0$.

\begin{figure}[bth]
\centering
\includegraphics[scale=0.7]{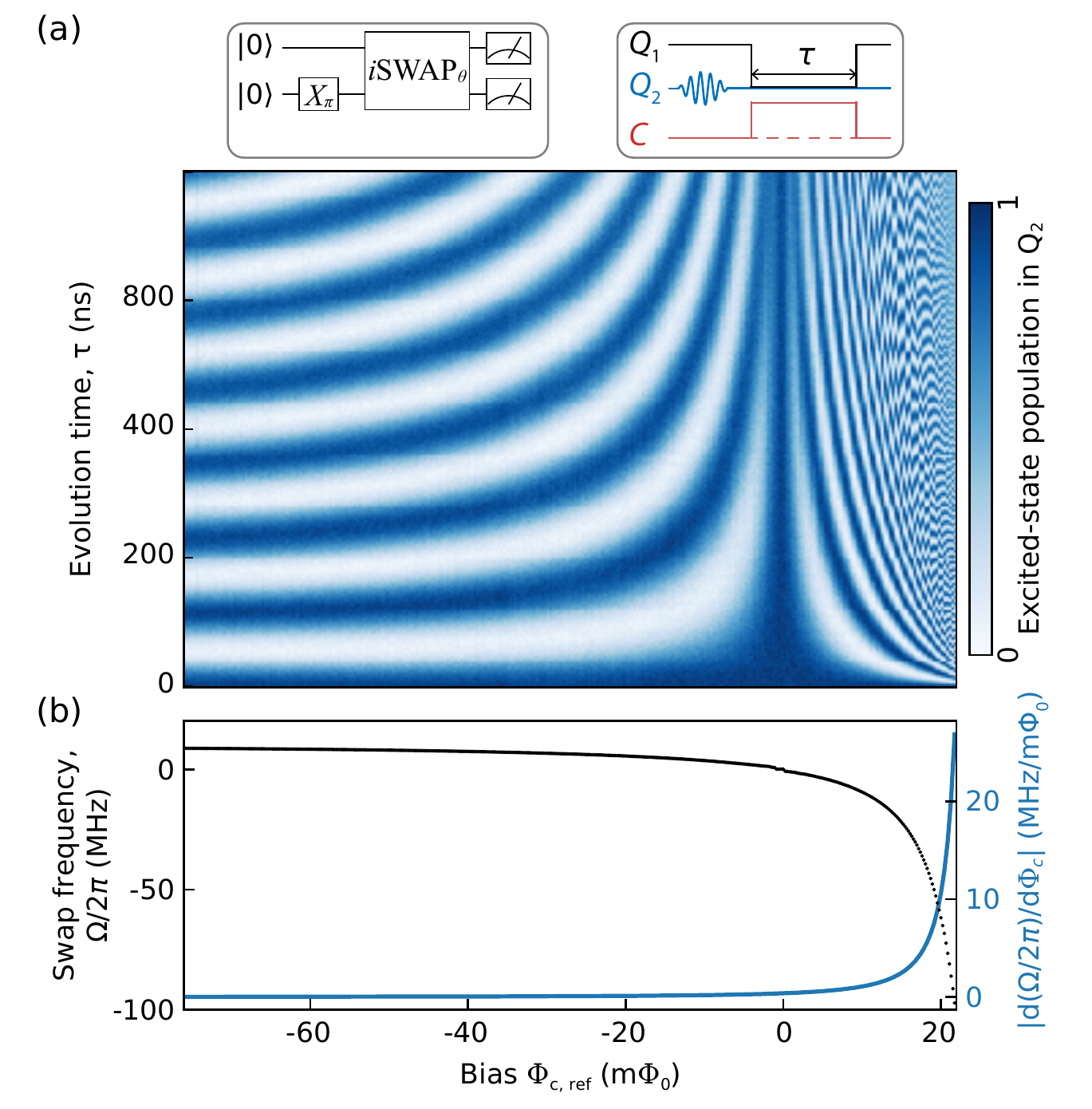}
\caption{Demonstration of tunable coupling. 
(a) By preparing the system in the $\ket{01}$ state and tuning $Q_1$ into a resonance with $Q_2$, we measured the excited-state population in $Q_2$ at different swapping times $\tau$ and coupler biases $\Phi_\mathrm{c}$. The gate sequence and corresponding waveforms are shown at the top of the panel. The variable swap operation is indicated by an $i\mathrm{SWAP}_\theta$ gate, where $\theta=\Omega\,\tau$ is the rotated angle. (b) From the frequencies of the fringes, we obtained the $\Omega(\Phi_\mathrm{c})$ dependence, tunable from +8.7~MHz to $-$97.1~MHz in the displayed range. The sign corresponds to the theoretical model defined in~\cite{yan_tunable_2018}. The coupling is completely turned off at the zero reference bias, i.e., $\Phi_\mathrm{c,ref}=0$. Its sensitivity to the flux, $\dd(\Omega/2\pi)/\dd\Phi_\mathrm{c}$, is obtained by taking the derivative.}
\label{fig2}
\end{figure}

To characterize the tunable coupling, we initialized the system to the state $\ket{01}$, set the pulse-bias of $Q_1$ to make it resonant with $Q_2$, pulse-biased the coupler from an idling point to a certain $\Phi_\mathrm{c}$, and waited for a varying duration before measuring the final population. The data are shown in Fig.~\ref{fig2}(a) at different flux biases $\Phi_\mathrm{c,ref}$ ($\Phi_\mathrm{c}$ referenced to the idling point). The fringing pattern describes the swapping dynamics between $\ket{10}$ and $\ket{01}$ with the oscillation frequency defined by $\Omega(\Phi_\mathrm{c,ref})$. Notably, at $\Phi_\mathrm{c,ref}=0$ (i.e.\ $\Phi_\mathrm{c}\approx0.17~\Phi_0$), the oscillations disappear, indicating that the coupling is turned off completely (we chose this bias as the idling point). The idling frequency of $Q_1$ was chosen to be approximately 35~MHz above $\omega_2$. 

From the oscillation frequency, we obtained the $\Omega(\Phi_\mathrm{c,ref})$ dependence, which, in good agreement with theory, varies smoothly from +8.7~MHz to $-$97.1~MHz over the measured range shown in Fig.~\ref{fig2}(b). The net coupling is dominated by direct qubit--qubit coupling (positive) when the coupler frequency is far above the qubit frequency ($\Phi_\mathrm{c,ref}<0$) and by coupler-mediated virtual exchange coupling (negative) when the coupler frequency is near the qubit frequency ($\Phi_\mathrm{c,ref}>0$). At a critical value $\omega_\mathrm{c}(\Phi_\mathrm{c,ref}\!=\!0)=6.15$~GHz, these two effects cancel out, leading to net zero coupling. The large tunability in the coupling enables fast two-qubit operations while eliminating residual coupling, which is necessary for the following experiment. Note that, in Fig.~\ref{fig2}(b), the derivative of $\Omega$ to $(\Phi_\mathrm{c,ref})$ increases dramatically when the coupler is biased into the negative-$\Omega$ regime, indicating strong sensitivity to flux fluctuations when the qubits are interacting with each other in this regime. 

For better understanding, we rewrite Eq.~\ref{hamiltonian_1} in a spin-1/2 Hamiltonian form,
\begin{equation}
    \mathcal{H} = \frac{1}{2} \left( \Omega(\Phi_\mathrm{c}) + \delta\Omega(\Phi_\mathrm{c}) \right)\, \sigma_Z + \frac{1}{2} \left( \Delta(\Phi_1) + \delta\Delta(\Phi_1) \right)\, \sigma_X,
    \label{hamiltonian_2}
\end{equation}
where $\sigma_{X,Y,Z}$ are the Pauli matrices. In this representation, the basis states are Bell states $\ket{10}\pm\ket{01}$ defined by the coupling term, so that we call it the $g$-frame. Accordingly, the qubit--qubit coupling ($\Omega$) is the $Z$-field, while the qubit--qubit detuning ($\Delta$) is the $X$-field. Because these two parameters can be independently controlled (after crosstalk corrections) by local flux lines, it is possible to perform arbitrary rotation to this subspace two-level system, the $g$-frame qubit. Additional properties of the $g$-frame qubit are summarized in Table~\ref{tab1} in comparison to a regular qubit.  $\delta\Omega(\Phi_\mathrm{c})$ and $\delta\Delta(\Phi_1)$ denote the fluctuations in the corresponding terms. In this device, flux noise is the dominant source for both $Z$ and $X$ noise because both $Q_1$ and the coupler are operated at flux-sensitive points.

Slow fluctuations in $\Omega$ cause diffusion of the angles during the swapping operation, causing the fringe amplitude to decay over time. It is easier to understand the dynamics by making an analogy with the conventional pure-dephasing phenomenon in a Ramsey or free-induction experiment~\cite{ramsey_molecular_1950}, where fluctuations in the qubit frequency desynchronize the phase coherence. Despite these similarities, a deviation from the targeted swap angle is a coherent two-qubit error. Passive methods, such as DD techniques, therefore have particular importance here.

\begin{figure}[bth]
\centering
\includegraphics[scale=0.5]{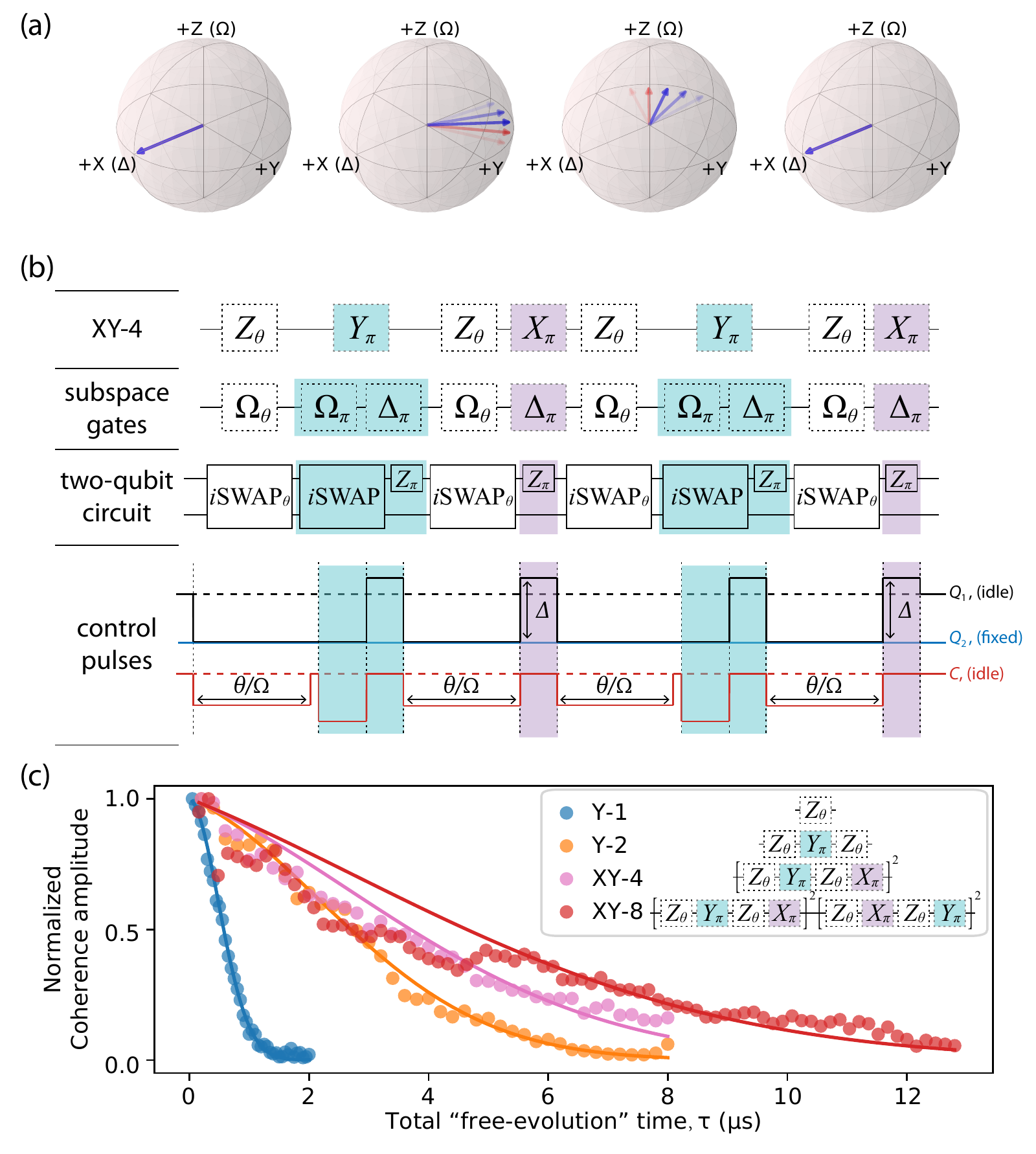}
\caption{Dynamical decoupling of qubit interactions. 
(a) Bloch picture of spin-echo evolution for the $g$-frame qubit.
The Cartesian coordinates are labelled $\Delta$, $Y$, and $\Omega$, indicating the corresponding control parameters.
In the second Bloch sphere, the fan-out of the Bloch vectors is a consequence of the $\Omega$-field inhomogeneities, a process analogous to single-qubit pure dephasing. After the refocusing pulse (here, an $X_\pi$ pulse), the vectors converge after the same amount of evolution time.
(b) Step-by-step compilation of the $XY$-4 sequence. 
Gate sequence under the single-qubit $XY$ definition (Row 1) is first translated to the $\Delta$--$Y$--$\Omega$ convention (Row 2); then, in the two-qubit quantum circuit (Row 3), the $\Delta$-rotation is realized by a single-qubit $Z$-rotation on $Q_1$ and the $\Omega$-rotation is realized by an $i\mathrm{SWAP}$-like operation between the two qubits; and finally, the actual control pulses are applied to the qubits and the coupler (Row 4). All pulses are nominally square pulses. The $\Omega_\pi$ gate ($\Omega=-25$~MHz, $\tau_\mathrm{gate}=20$~ns) and the $\Delta_\pi$ gate ($\Delta=83$~MHz, $\tau_\mathrm{gate}=6$~ns) are pre-calibrated using pulse-train techniques (see \cite{supplement} for details). The $\Omega_\theta$ ($i\mathrm{SWAP}_\theta$) gate acts as an always-on free-evolution unitary, which is an $i\mathrm{SWAP}$-like operation with a variable swapping rate $\Omega$ and a duration $\tau_\theta=\theta/\Omega$. In the experiment we chose the $YXYX$ order, but there is no nominal difference from the $XYXY$ order.
(c) Measured coherence decay traces at $\Phi_\mathrm{c,ref} = 18.5$~m$\Phi_0$ for various dynamical-decoupling (DD) sequences, including $Y$-1, $Y$-2, $XY$-4, and $XY$-8. Solid lines are the decay in the form of $f(t)=A e^{-\Bar{\Gamma}_1 t-(\Gamma_{\varphi} t)^2}+B$ after correcting for the beating pattern caused by the $\delta\Delta$ noise (see \cite{supplement} for details). The inset shows the gate sequences under the $XY$ definition, which can be compiled according to the same rule as shown in panel (b).}
\label{fig3}
\end{figure}

To understand how DD techniques work for the $g$-frame qubit, we can first look at the state evolution under the spin-echo~\cite{hahn_spin_1950} protocol with a single refocusing pulse, as shown by the Bloch-picture dynamics in Fig.~\ref{fig3}(a).
After initializing the state to $\ket{10}$ or $\ket{01}$, the system state is within the subspace (first Bloch sphere). The state of the $g$-frame qubit will precess around the longitudinal ($Z$) axis at a fluctuating rate $\Omega(\Phi_\mathrm{c}) + \delta\Omega(\Phi_\mathrm{c})$. Given the usually low-frequency nature of flux noise, state vectors in different realizations precess at non-uniform rates and gradually dephase (second Bloch sphere). At a certain timepoint, an $X$ pulse, or a $180^\circ$ $\Delta$-rotation flips these vectors (third Bloch sphere), which then continue to precess at the same rate as before, such that they converge again after the same amount of evolution time as before the $X$ pulse (fourth Bloch sphere). It can be seen that noise with correlation times longer than the free-precession period can be effectively suppressed.

In the experiment, we performed the spin-echo protocol and its generalization, periodic DD sequences~\cite{carr_effects_1954,meiboom_modified_1958}, in which the free evolution is split into $N$ equal sections by an array of refocusing pulses, e.g., $N=1$ for Ramsey and $N=2$ for spin echo. We also performed various types of pulse trains, including all-$X$, all-$Y$, and alternating $XY$ pulses~\cite{gullion_new_1990}. Figure \ref{fig3}(b) illustrates the compilation steps of the $XY$-4 sequence ($XY$ denotes the pulse type and 4 denotes the number of split periods). In general, an $X$-rotation is translated to a $\Delta$-rotation, while a $Z$-rotation is translated to an $\Omega$-rotation. The $Y_\pi$ gate is implemented with a composite pulse, i.e., a $Z_\pi$ gate followed by an $X_\pi$ gate. 

Figure \ref{fig3}(c) shows a few selected decay traces, measured at $\Phi_\mathrm{c,eff}=18.5$~m$\Phi_0$, which is in the $\delta\Omega$-dominant regime. The results show progressive coherence improvement with more advanced sequences.
Note that the displayed echo trace was taken with a $Y_\pi$ pulse. Because the condition of a good refocusing pulse is a $\pi$-rotation around an axis transverse to the noise axis~\cite{borneman_application_2010}, the sequence with $X_\pi$ pulses is sensitive to $\delta\Delta$ fluctuations, while the sequence with a $Y_\pi$ pulse is robust against both $\delta\Omega$ and $\delta\Delta$ fluctuations and is therefore a preferred choice.
To further decouple the system from the noise in all directions, we used a mixture of $X$ and $Y$ gates such as the $XY$-4 and $XY$-8 sequences~\cite{tyryshkin_dynamical_2010,souza_robust_2012}.
We find that the decay traces show stronger beating patterns with increasing $N$, as a consequence of the $\delta\Delta$ fluctuations which adds complication to the dynamics by tilting the rotation axis (see \cite{supplement} for details). 

\begin{figure}[bth]
\centering
\includegraphics[scale=0.6]{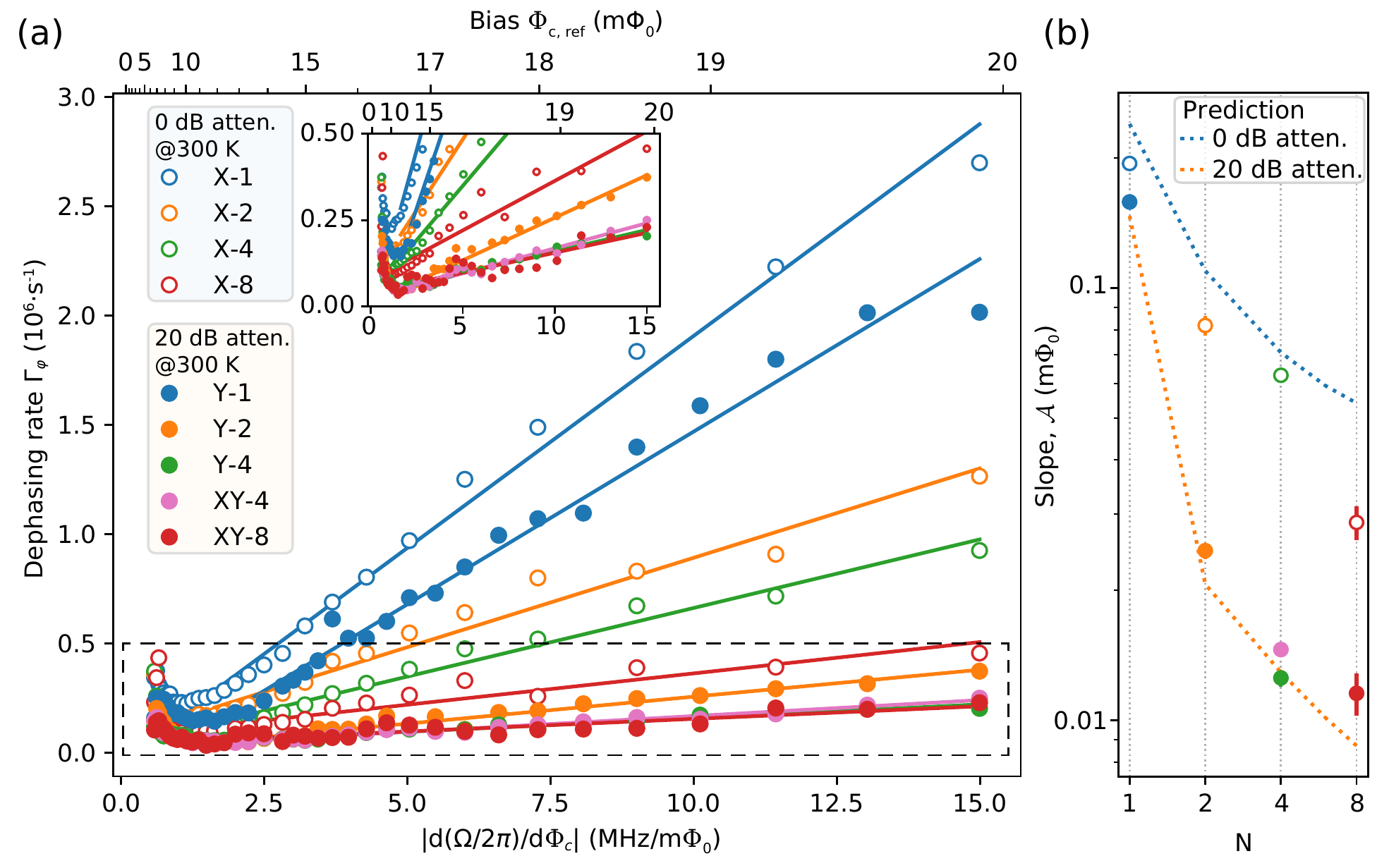}
\caption{Noise characterization. 
(a) Extracted pure-dephasing rate $\Gamma_{\varphi}$ versus the flux sensitivity $\dd(\omega/2\pi)/\dd\Phi_\mathrm{c}$ (bottom axis) or the corresponding flux biases (top axis) obtained from Fig.~\ref{fig2}(b) for various DD sequences. Two different setups of room-temperature attenuation were used in these sequences. Note that $X$-1 and $Y$-1 are the same sequence (Ramsey). Solid lines are linear fits in the $\Phi_\mathrm{c}$-sensitivity regime, $\dd(\Omega/2\pi)/\dd\Phi_\mathrm{c} \geq 1.5~\mathrm{MHz/m\Phi_0}$. The inset shows a magnification of the dashed area. The extracted linear slope is plotted in panel (b) for each sequence type (denoted by the number of free-induction periods $N$). The dashed line indicates the values estimated from the combination of directly measured signal generator noise and inferred ground-loop noise using the filter-function formalism.}
\label{fig4}
\end{figure}

To disentangle the influence from the $\delta\Delta$ noise, we performed the same set of DD sequences at different flux biases. According to the filter-function formalism, the (Gaussian) pure-dephasing rate can be expressed as
\begin{equation}
    \Gamma_\varphi = \left| \frac{\dd\Omega}{\dd\Phi_\mathrm{c}} \right| \sqrt{ \frac{1}{2\pi}\int_0^\infty{\dd\omega \, S_{\Phi_\mathrm{c}}(\omega) F(\omega,\tau^*)} } = \left| \frac{\dd\Omega}{\dd\Phi_\mathrm{c}} \right| \mathcal{A}(N),
    \label{dephasing}
\end{equation}
where $S_{\Phi_\mathrm{c}}(\omega)$ is the noise spectral density and $F(\omega,\tau^*)$ is a filter function unique to the applied DD sequence with a characteristic total duration $\tau^*$. This integral is not sensitive to the choice of $\tau^*$ in the current case (see \cite{supplement} for details).
Therefore, given certain $S_{\Phi_\mathrm{c}}(\omega)$ and $\tau^*$, the square-root term can be treated as a function of $N$ ($\mathcal{A}(N)$), which depends only on the type of the sequence used.
As expected from Eq.~\ref{dephasing}, the measured pure-dephasing rate is proportional to the noise sensitivity $\dd\Omega/\dd\Phi_\mathrm{c}$ in the $\delta\Phi_\mathrm{c}$-dominant regime, as shown in Fig.~\ref{fig4}(a). 
The extracted slopes from linear fits are plotted in Fig.~\ref{fig4}(b) for various DD sequences and are compared with the numerically calculated $\mathcal{A}(N)$. A few findings regarding the performance of the DD sequences and the noise model of our system are discussed below.

To investigate the $\delta\Phi_\mathrm{c}$ noise sources, we tried two different attenuation setups, 0~dB and 20~dB, at room temperature over the $Z$ control lines. As seen in Fig.~\ref{fig4}(b), the swap-dephasing rates with the additional 20-dB attenuation are drastically smaller for $N\geq2$. By using the filter-function formalism, we found that the abnormally strong noise in the no-attenuation case can be well explained by noise measured directly from the pulse generator. This noise is effectively suppressed by simply adding attenuation while exploiting the output range of the generator. 

In the case of added 20-dB attenuation, the $XY$-8 sequence generally has improved the pure-dephasing time by $\sim$14 times, compared to the $Y$-1 sequence, which demonstrates the effectiveness of using DD sequences to suppress noise in two-qubit interaction. In addition, we noticed that the extra attenuation shows little improvement in the $N=1$ (Ramsey) case. We suspect the reason is extra low-frequency noise from ground loops which cannot be attenuated (see \cite{supplement} for details).
Using a model combining both instrumentation and ground-loop noise, we find good agreement with the DD results shown in Fig.~\ref{fig4}(b).

So far, analyses have focused on the $\delta\Phi_\mathrm{c}$-sensitive regime. In this regime, the strong quantization field, $\Omega$, effectively suppresses swap-dephasing from $\delta\Delta$ or $\delta\Phi_\mathrm{1}$ noise to the second order. In the low-$\Omega$ regime, $\delta\Delta$ becomes the dominant noise source and its influence is augmented by small $\Omega$. This explains the tilt-up of the dephasing rates towards diminishing $\Omega$ (smaller $\Phi_\mathrm{c,eff}$).

To conclude, we demonstrated a compact tunable-coupler design with a large dynamic range in a superconducting quantum circuit. Taking advantage of the full controllability over the one-excitation subspace of a two-qubit system, we implemented $g$-frame analogues of DD sequences and demonstrated effective suppression of swap-dephasing, a type of coherent two-qubit error. The results suggest the importance of including such errors in the error model when developing fault-tolerant quantum computers. Our demonstration introduces the open-loop quantum control technique to the realm of two-qubit interactions, an important first step towards non-local error mitigation. Our work also establishes the framework for controlling the $g$-frame qubit, which can be the foundation for developing dynamically corrected two-qubit gates~\cite{khodjasteh_dynamically_2009} to further enhance system performance at large scales.

\section*{Acknowledgements}
We thank Simon Gustavsson, Xiaotong Ni, Youngkyu Sung for insightful discussions, and Martha Evonuk for assistance in editing. This work was supported by the Key-Area Research and Development Program of Guang-Dong Province (Grant No. 2018B030326001), the National Natural Science Foundation of China (U1801661), the Guangdong Innovative and Entrepreneurial Research Team Program (2016ZT06D348), the Guangdong Provincial Key Laboratory (Grant No.2019B121203002), the Natural Science Foundation of Guangdong Province (2017B030308003), and the Science, Technology and Innovation Commission of Shenzhen Municipality (JCYJ20170412152620376, KYTDPT20181011104202253), and the NSF of Beijing (Grants No. Z190012).

\bibliographystyle{naturemag}
\bibliography{references}

\newpage

\begin{table}[]
    \centering
    \caption{Comparison between a single-qubit system and a $g$-frame qubit encoded in the one-excitation subspace of two qubits.
    Interestingly, for the $g$-frame qubit, the observed relaxation rate $\Gamma_{1g}$ depends not only on the averaged relaxation rates of both qubits $\Bar{\Gamma}_1$, i.e., the rate at which the population leaks out of the subspace, but also on $\Gamma_\Omega$, the bit-flip rate between the basis states due to the noise transverse to the eigenaxis (i.e., $\delta\Delta$ noise) at the eigenfrequency ($\Omega$). }
    \begin{tabular}{lll}
        \hline

        \ 
        & Laboratory-frame qubit & $g$-frame qubit ($\Delta=0$) \\

        \hline

        Hamiltonian
        & $\mathcal{H} = \frac{1}{2}\omega_q\sigma_Z$  & $\mathcal{H} = \frac{1}{2}\Omega\sigma_Z$  \\
        
        Basis states
        & $\ket{0}$, $\ket{1}$  & $\frac{1}{\sqrt2}\left(\ket{10}-\ket{01}\right)$, $\frac{1}{\sqrt2}\left(\ket{10}+\ket{01}\right)$  \\

        Qubit operation
        & \   & \  \\

        \ \ Z-rotation $Z_\theta$
        & $\omega_\mathrm{q}(t)\sigma_Z$ & $\Omega(t)\sigma_Z$ \\

        \ \ X-rotation $X_\theta$
        & $A(t)\cos(\omega_\mathrm{q} t)\sigma_X$ & $\Delta(t)\sigma_X$ \\

        \ \ Y-rotation $Y_\theta$
        & $A(t)\cos(\omega_\mathrm{q} t + \frac{\pi}{2})\sigma_X$ & $X_{-\frac{\pi}{2}} Z_\theta X_\frac{\pi}{2}$, in particular $Y_\pi = Z_\pi X_\pi$ \\

        Relaxation
        & \   & \  \\

        \ \ Decay law
        & $\exp\left[-\Gamma_1\tau\right]$ & $\exp\left[-\Gamma_{1g}\tau\right]$  \\

        \ \ Noise source
        & $\Gamma_1 = \frac{1}{2} S_{\perp Z}(\omega_q)$ & $\Gamma_{1g} = \Bar{\Gamma}_1 + \Gamma_\Omega$, where \\
        & & $\Bar{\Gamma}_1 = \frac{1}{2} (\Gamma_{1,Q_1} + \Gamma_{1,Q_2})$, $\Gamma_\Omega = \frac{1}{2} S_\Delta(\Omega)$ \\
        
        Dephasing (Gaussian)
        & \   & \  \\
        
        \ \ Decay law
        & $\exp\left[-\frac{1}{2} \Gamma_1\tau - (\Gamma_\varphi\tau)^2\right]$ & $\exp\left[-(\Bar{\Gamma}_1 + \frac{1}{2} \Gamma_\Omega)\tau - \left(\Gamma_{\varphi g}\tau\right)^2\right]$  \\

        \ \ Noise source
        & $S_Z(\omega)$ & $S_\Omega(\omega)$, and $S_\Delta(\omega)$ when $\Omega$ is small  \\

        \hline
    \end{tabular}
    \label{tab1}
\end{table}

\end{document}

\title{Supplementary Material for Suppressing Coherent Two-Qubit Errors via Dynamical Decoupling}

\author{Jiawei Qiu}
\thanks{These authors contributed equally.}
\affiliation{\siqse}\affiliation{\physustech}
\author{Yuxuan Zhou}
\thanks{These authors contributed equally.}
\affiliation{\siqse}\affiliation{\physustech}

\author{Chang-Kang Hu}
\affiliation{\siqse}\affiliation{\gdpkl}\affiliation{\szkl}
\author{Jiahao Yuan}
\affiliation{\siqse}\affiliation{\physustech}
\author{Libo Zhang}
\affiliation{\siqse}\affiliation{\gdpkl}\affiliation{\szkl}

\author{Ji Chu}
\affiliation{\siqse}
\author{Wenhui Huang}
\affiliation{\siqse}\affiliation{\physustech}
\author{Weiyang Liu}
\affiliation{\siqse}\affiliation{\gdpkl}\affiliation{\szkl}
\author{Kai Luo}
\affiliation{\siqse}\affiliation{\physustech}
\author{Zhongchu Ni}
\affiliation{\siqse}\affiliation{\physustech}
\author{Xianchuang Pan}
\affiliation{\siqse}
\author{Zhixuan Yang}
\affiliation{\siqse}
\author{Yimeng Zhang}
\affiliation{\siqse}

\author{Yuanzhen Chen}
\affiliation{\siqse}\affiliation{\physustech}\affiliation{\gdpkl}\affiliation{\szkl}
\author{Xiu-Hao Deng}
\affiliation{\siqse}\affiliation{\gdpkl}\affiliation{\szkl}
\author{Ling Hu}
\affiliation{\siqse}\affiliation{\gdpkl}\affiliation{\szkl}
\author{Jian Li}
\affiliation{\siqse}\affiliation{\gdpkl}\affiliation{\szkl}
\author{Jingjing Niu}
\affiliation{\siqse}\affiliation{\gdpkl}\affiliation{\szkl}
\author{Yuan Xu}
\affiliation{\siqse}\affiliation{\gdpkl}\affiliation{\szkl}
\author{Tongxing Yan}
\affiliation{\siqse}\affiliation{\gdpkl}\affiliation{\szkl}
\author{Youpeng Zhong}
\affiliation{\siqse}\affiliation{\gdpkl}\affiliation{\szkl}
\author{Song Liu}
\thanks{lius3@sustech.edu.cn}
\affiliation{\siqse}\affiliation{\gdpkl}\affiliation{\szkl}
\author{Fei Yan}
\thanks{yanf7@sustech.edu.cn}
\affiliation{\siqse}\affiliation{\gdpkl}\affiliation{\szkl}
\author{Dapeng Yu}
\affiliation{\siqse}\affiliation{\physustech}\affiliation{\gdpkl}\affiliation{\szkl}

\maketitle

\section{Device Design and Fabrication}
\begin{figure}[bth]
    \centering
    \includegraphics[scale=0.7]{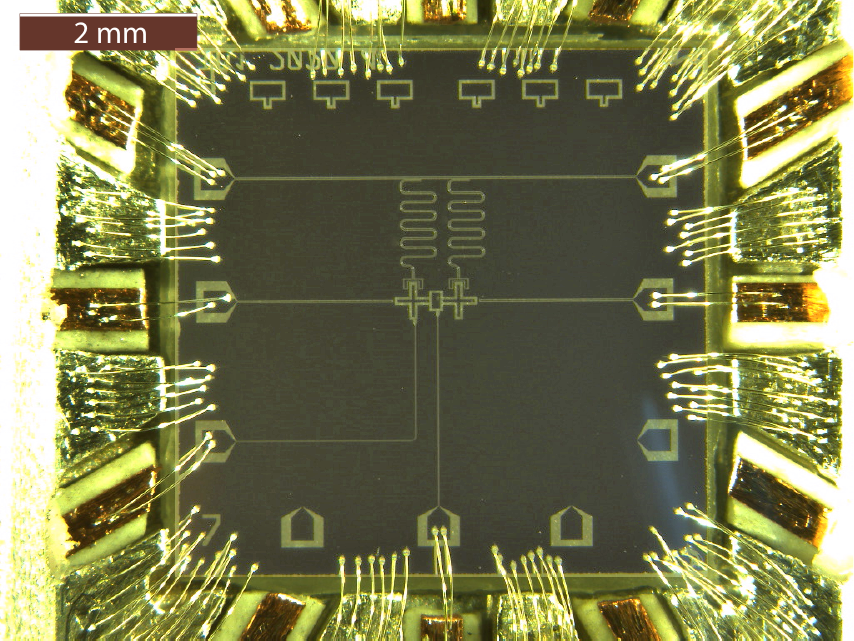}
    \caption{
        Micorscope of the mounted chip.
    }
\label{fig:mounted}
\end{figure}

The device has two Xmons~\cite{barends_coherent_2013} and one transmon qubit~\cite{koch_charge_2007}. The layout can be found in maintext or Fig.\ref{fig:mounted}. The two Xmons have individual controls and are individually read out. The left Xmon is made tunable so it has both wave and frequency controls, while the right one is fixed-frequency with only RF control for it. The tramsmon in the mid serves as tunable coupler and was kept on ground state. It also came with DC SQUID so it can be tuned by the center bias line.

The tunable coupling scheme is implemented as \cite{yan_tunable_2018} decribed. Xmons are coupled vis two channels: direct capacitive coupling and indirect coupling mediated by coupler. The net effective qubit-qubit coupling strength is given by $ \widetilde{g} \cong g_1g_2/\Delta+g_{12}$
where $g_j (j=1,2) $ is the coupling strength between qubits ($\omega_1$ and $\omega_2$) and coupler ($\omega_c$),  $g_{12}$ is the direct coupling strength of two qubits. Note that $ 1/\Delta=[1/(\omega_1-\omega_c)+1/(\omega_2-\omega_c)]/2<0$, hence the strength of the net coupling can be adjusted by changing the coupler frequency, at certain flux bias the net coupling $\tilde{g}$ can tuned to 0. In order to making the zero-coupling point accessible, one has to carefully choose parameters among the circuit, and hence the coupler is finally decided to be of rectangular shape as shown in Fig.1(b). Simulation shown the capacitance of qubits and coupler are $C_1=C_2=81.4fF$ and $C_c=44.4fF$ respectively.

The device were fabricated with a two-step process. All the circuit geometry except junctions are defined within one time of photolithography, with roughly 100nm-thick Al thin film deposited on a c-plane sapphire substrate in a PLASSYS system with a growth rate of 1 nm/s. Etching is done with a inductively coupled plasma (ICP) dry etcher with $\mathrm{BCl}_3$/$\mathrm{Cl}_2$. Josephson junctions are defined by double-angle evaporation of aluminum in a high vacuum electron-beam evaporator. Junctions were patterned using the bridge-free ‘Manhattan Style’ by e-beam lithography for precise control over qubit frequencies. The junctions were realized with two separate angle-evaporated aluminum layers, with a layer of oxide in between. The oxide is introduced by in-situ oxidation with oxygen at 3.3 torr for about 20 minutes. In order to restrain the loss due to argon milling on the substrate, we deposit another Al film on top of the Josephson junction and connecting leads as additional galvanic contact~\cite{dunsworth_characterization_2017}.

\begin{figure}[bth]
    \centering
    \includegraphics[scale=0.7]{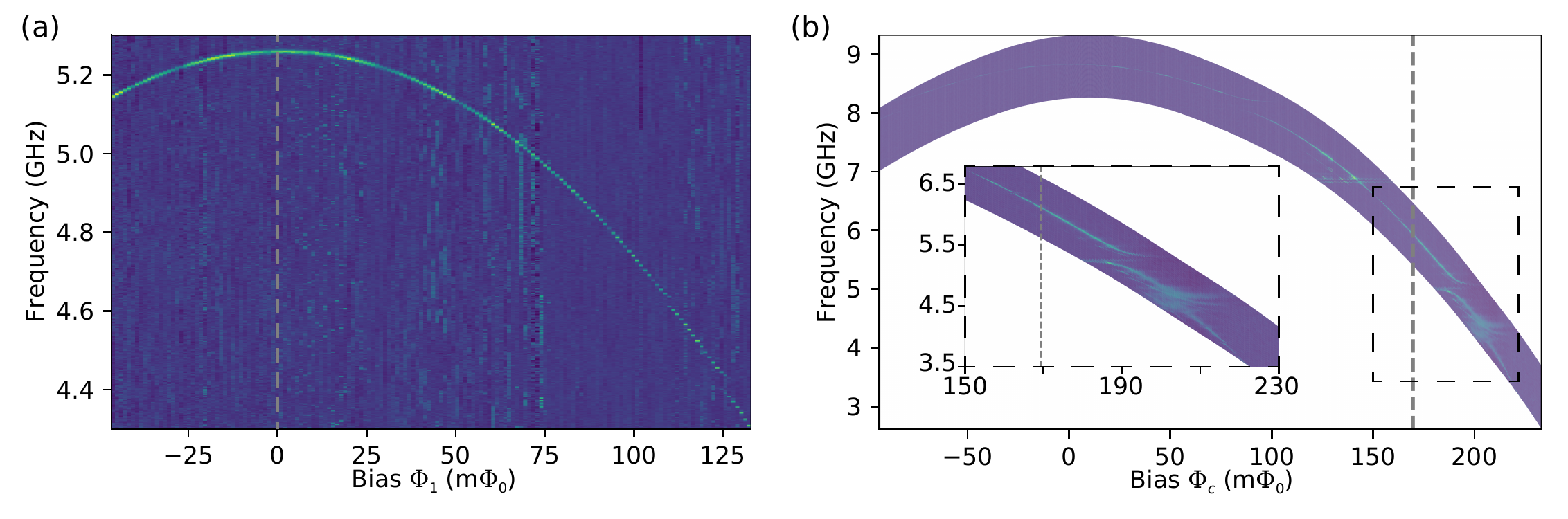}
    \caption{
        The frequency spectrum of (a). The tunable qubit $Q_1$ and (b). Coupler $C$. Inset: zoom in for the area for our experiment. The gray dashed line marks the idling bias of coupler.
    }
\label{fig:spec}
\end{figure}

The calibrated sample parameters and shown in Table.\ref{tab:device_parameters}. Note that the coherence performance of $Q_1$ varies in different electronics configurations, we will come back to this later. The frequency spectrum of $Q_1$ and coupler is shown in Fig:\ref{fig:spec}, with standard spectrum scan. Readout of coupler is via the cavity of $Q_2$. The idling bias point, or the reference bias is indicated with dashed line in the figure.

\begin{table*}[]
    \centering
    \begin{tabular}{c|c|c}

        \hline

        Parameters
        & $Q_1$ & $Q_2$ \\

        \hline

        Resonator frequency ($\omega_\mathrm{R1}/2\pi$, $\omega_\mathrm{R2}/2\pi$)
        & 6.955 GHz & 7.001 GHz  \\
        
        Qubit frequency ($\omega_{1, \Phi_1=0}/2\pi$, $\omega_2/2\pi$)
        & 5.270 GHz  & 4.614 GHz  \\

        Qubit anharmonicity ($\alpha_1/2\pi$, $\alpha_2/2\pi$)
        & $-210$ MHz  & $-240$ MHz \\
        
        Coupler frequency ($\omega_\mathrm{c,~idling}/2\pi$)
        & \multicolumn{2}{c}{6.150 GHz} \\
        
        Coupler anharmonicity ($\alpha_\mathrm{c}/2\pi$)
        & \multicolumn{2}{c}{$-370$ MHz} \\
        
        Qubit-coupler coupling ($g_\mathrm{1c}/2\pi$, $g_\mathrm{2c}/2\pi$)
        & 122 MHz & 105 MHz \\
        
        Qubit-qubit direct coupling ($g_{12}/2\pi$)
        & \multicolumn{2}{c}{12 MHz} \\

        Qubit relaxation time ($T_1$)
        & 9.68 $\mu$s & 12.78 $\mu$s \\

        Qubit dephasing rate ($T_{\phi, \mathrm{Ramsey}}$)
        & 0.76 $\mu$s & 29.28 $\mu$s \\

        Qubit dephasing rate ($T_{\phi, \mathrm{echo}}$)
        & 3.82 $\mu$s & 125.7 $\mu$s \\

        \hline

    \end{tabular}
    \caption{Parameters of sample in our experiment.}
    \label{tab:device_parameters}
\end{table*}

\section{Measurement Setup}
\begin{figure}[bth]
    \centering
    \includegraphics[scale=0.7]{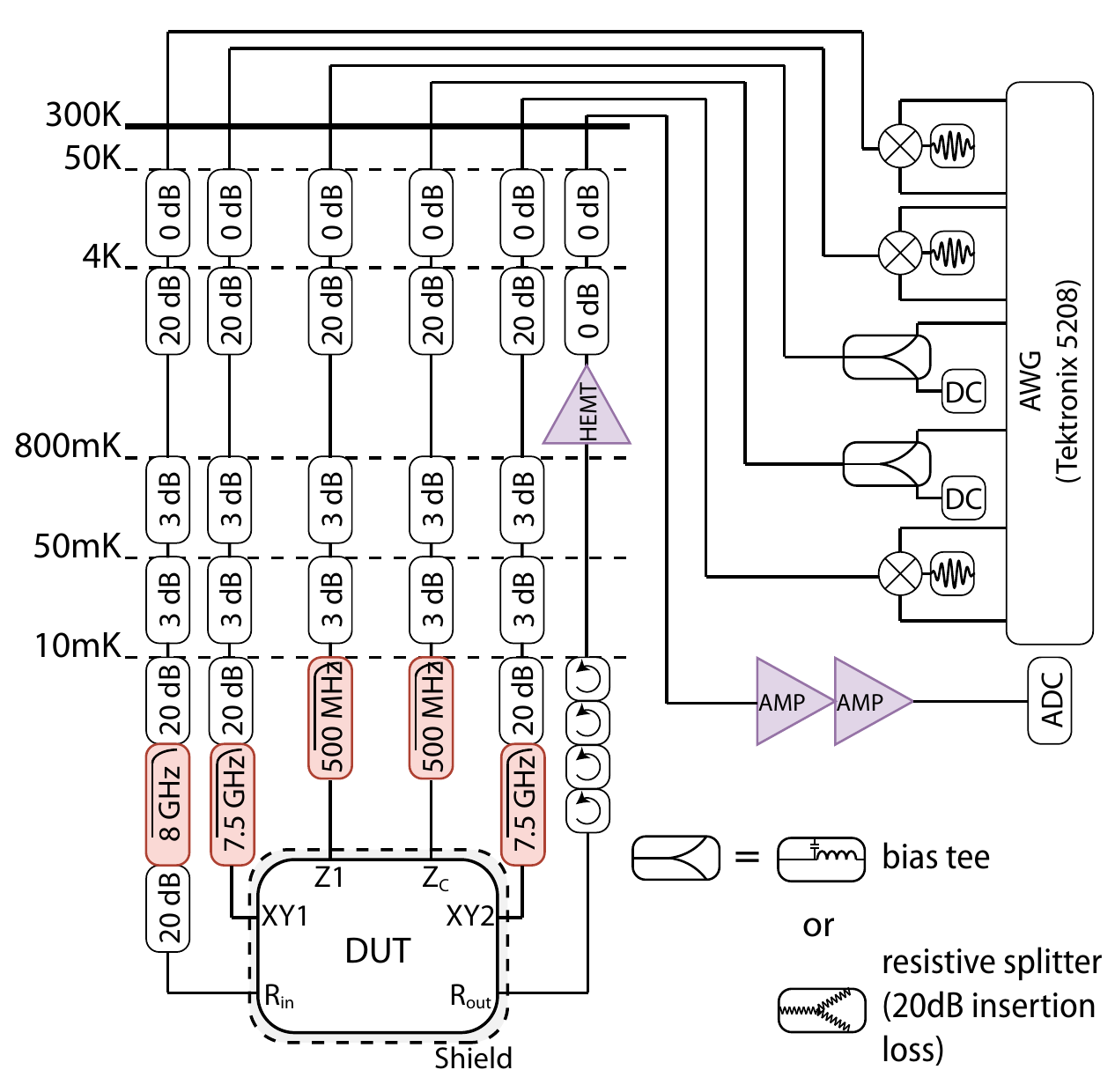}
    \caption{
        Schemetic diagram of wirings and circuit components. The combiner for DC signal and flux pulse is either a bias tee or a resistive splitter.
    }
\label{fig:wiring}
\end{figure}

The device is mounted in an aluminum box and the cooled down to 10mK in a dilution refrigerator. A additional magnetic shield is used to protect qubits from external flux fluctuations. Wiring and control electronics is shown in Fig.\ref{fig:wiring}.

Note that the two different configurations are used in our experiment: one use bias tee to combine the flux bias pulse and DC offset signal, the other use a resistive splitter as combiner. The RF port of the bias comes with a high pass property with cutoff freuqency at 100 kHz, while the resistive splitter introduces an additional 20 dB insertion loss on signal. In the later configuration, we observed appearant longer decoherence time on both $Q_1$ and $g$-frame qubit, as discussed in maintext. Most of the noise comes from the pulse generator (Tektronics 5208 AWG). Besides we use other instruments to implement control on our system: we use a Keysight MXG-N5183B signal generator for local oscillator, a Rohde\&Schwarz SGS100A the synthesize the RF control pulses, we use a Rigol 1062 for DC signal and an ATS digitizer for readout signal measurement. The amplifiers at room temperature is MITEQ AFS3-00101200-22-10P-4. All of them more or less introduces some noise.

\section{Qubit Dephasing} \label{sec:qubit dephasing}
\begin{figure}[bth]
    \centering
    \includegraphics[scale=0.7]{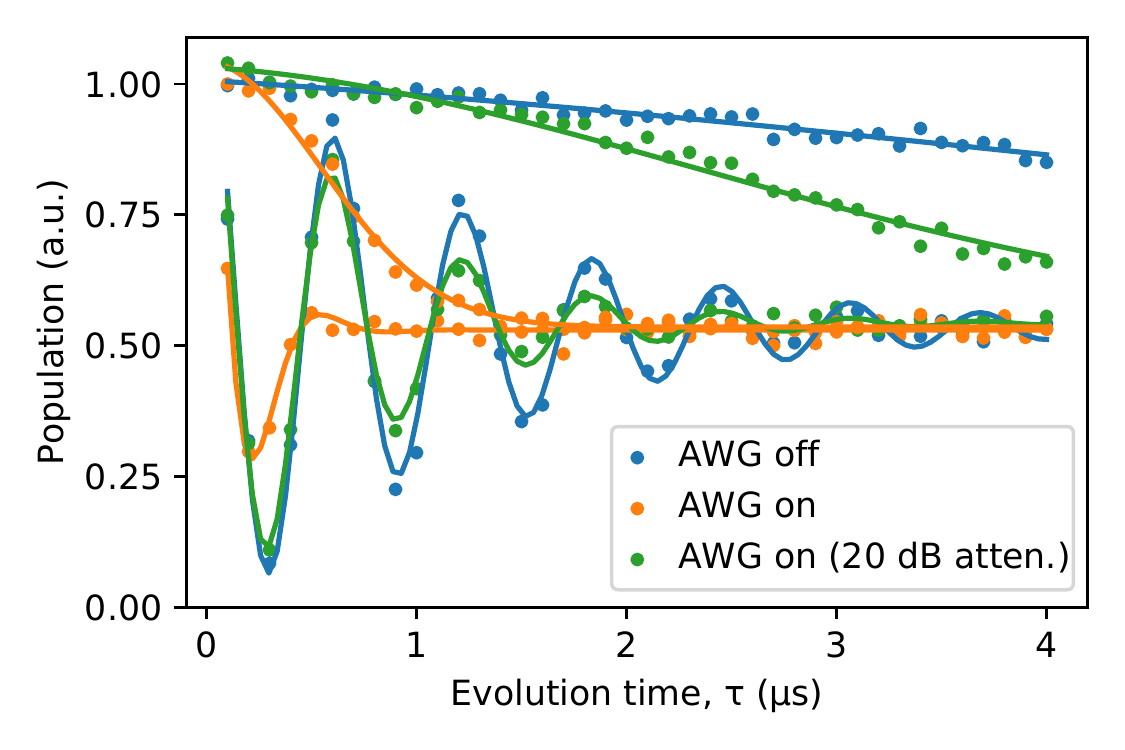}
    \caption{
        Decoherence of $Q_1$ in with different AWG configurations.
    }
\label{fig:q1tphi}
\end{figure}

In our experiment, $Q_1$ is biased around sweet point and decoheres in presence of flux noise. The measured $T_{\phi, \mathrm{Ramsey}}$ varies with different configurations of AWG off/on/attenuated with a 20dB attenuator on output port of AWG, as shown in Fig.\ref{fig:q1tphi}. The detailed data is in Table.\ref{tab:q1tphi}. Notice the dephasing rate with AWG on is 5.758$\times10^6\,\mathrm{s}^{-1}$ or 0.537$\times10^6\,\mathrm{s}^{-1}$ more than that with AWG off in case with or without attenuation, this number is 1.179$\times10^6\,\mathrm{s}^{-1}$ and 0.164$\times10^6\,\mathrm{s}^{-1}$ in echo experiment. the improvement by attenuator is about 10 in both cases. This indicates that at least two noise sources contributes to dephasing of $Q_1$: one is from the AWG and the other one cannot be attenuated by attenuator on AWG. These two parts have comparable intensity, and the later part is suppressed better with echo experiment, indicating more low-frequency components in it. Assuming the noise spectrum is with form $1/f^\alpha$, we infer that $\alpha\approx1.1$ with the ratio $\Gamma_{\phi, \mathrm{echo}} / \Gamma_{\phi, \mathrm{Ramsey}} \approx 7.9$ with AWG turning off.

\begin{table}[]
    \centering
    \begin{tabular}{ccc}

        \hline

        Config.
        & $\Gamma^1_{\phi, \mathrm{Ramsey}}$ ($10^6\,\mathrm{s}^{-1}$) & $\Gamma^{1}_{\phi, \mathrm{echo}}$ ($10^6\,\mathrm{s}^{-1}$)  \\

        \hline

        AWG off
        & 0.771 & 0.098  \\
        
        AWG on
        & 6.529  & 1.277  \\
        
        AWG on (20 dB atten.)
        & 1.308  & 0.262  \\

        \hline

    \end{tabular}
    \caption{Dephasing rate of $Q_1$ in different AWG configurations.}
    \label{tab:q1tphi}
\end{table}

\section{Calibrating Operations on $g$-frame Qubit}
Precise $\Omega_\pi$ and $\Delta_\pi$ operations are required to implement the dynamical decoupling (DD) sequences in $g$-frame. The operation are realized by pulse biasing on frequency of coupler or $Q_1$. The pulse is calibrated with "pulse train" technique: starting with a system in ground state $\ket{00}$, by apply a $X_\pi$ pulse on $Q_2$ we initialize the system with state $\ket{01}$. Then a number (usually even number) of $\Omega$ pulses are applied with spacing of 2ns before measurements. Applying multiple pulses is for better sensitivity of detecting pulse error. The experiment is repeated with sweeping pulse amplitude, while the pulse shape is square pulse with fixed width (20 ns for $\Omega$ pulse, 6 ns for $\Delta$ pulse). In case where evolution angle integrated to $\pi$, the final state of system should be $\ket{01}$ again, with measured population on $Q_2$ reaches a maximum.

\begin{figure}[bth]
    \centering
    \includegraphics[scale=0.7]{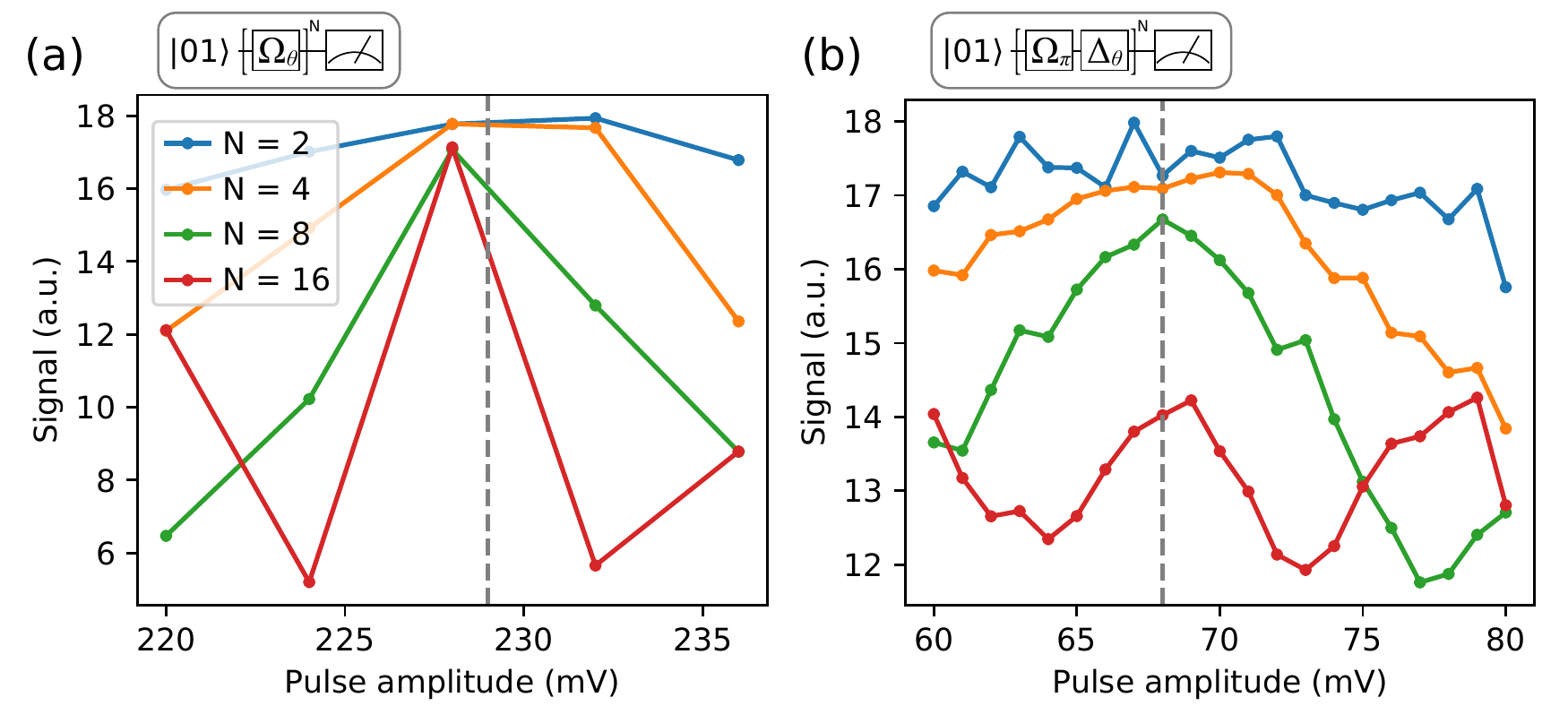}
    \caption{
        Result of calibrating operations in $g$-frame. (a). $\Omega_\pi$ gate. (b). $Y_\pi$ gate. The operation sequences are shown on the top.
    }
\label{fig:cal_ggate}
\end{figure}

However, $\Delta_\pi$ pulse cannot be calibrated like above, because the initial state $\ket{01}$ is eigenstate of $\Delta_\pi$. We turned to calibrate the $Y_\pi$ pulse, which is composited by $\Omega_\pi\times\Delta_\pi$, by sweeping pulse amplitude of the $\Delta$ pulse part. A good value for $\Delta_\pi$ pulse is then calibrated with a $Y_\pi$ operation. The result is shown at Fig.\ref{fig:cal_ggate}. The calibrated amplitudes are 229 mV and 68mV for $\Omega$ pulse and $\Delta$ pulse respectively.

\section{Decoherence of $g$-frame Qubit}
Recall the $g$-frame qubit is subjected to transverse ($\Delta$) noise and longitudinal ($\Omega$) noise:
\begin{equation}
    \mathcal{H} = \frac{1}{2} \left( \Omega(\Phi_\mathrm{c}) + \delta\Omega(\Phi_\mathrm{c}) \right)\, \sigma_Z + \frac{1}{2} \left( \delta\Delta(\Phi_1) \right)\, \sigma_X.
\end{equation}
Both of them contribute to decoherence of the $g$-frame qubit: Transverse noise $S_\Delta(\omega)$ leads to longitudinal depolarization with rate donated by $\Gamma_{\Omega}$. Longitudinal noise $S_\Omega(\omega)$ leads to transversal depolarization with rate donated by $\Gamma_{\varphi g}$. Besides, relaxation on each qubit site also makes the $g$-frame qubit decoheres. Providing weak coupling between qubits and environments and the noise process to be Markovian, we can obtain a exponential decay trace with Linbdlad master equation:
\begin{equation}
    \dot{\rho}(t)
    = \frac{1}{i\hbar}\left[\mathcal{H}, \rho(t)\right]
      + \sum_n{\frac{1}{2}\left[2L_n\rho(t)L_n^\dagger - \rho(t)L_n^\dagger L_n - L_n^\dagger L_n\rho(t) \right]},
\end{equation}
with the complete system hamiltonian $\mathcal{H} = \frac{\omega_1}{2}\sigma_Z\otimes I + \frac{\omega_2}{2}I\otimes\sigma_Z + \frac{\Omega}{2}(\sigma_-\otimes\sigma_+ + \sigma_+\otimes\sigma_-)$ and Linbdlad operators $\sqrt{\Gamma_{1,Q1}}(\sigma_-\otimes I)$, $\sqrt{\Gamma_{1,Q2}}(I\otimes \sigma_-)$, $\sqrt{\Gamma_\Omega/2}(\sigma_Z\otimes I)$ for relaxation of $Q_1$, $Q_2$ and $g$-frame qubit respectively. The pure dephasing rate process is subjected to noise with long correlation time and hence breaks the Markovian approximation. We will discuss it later. And only flux noise on $Q_1$ is taken into account for $\Gamma_\Omega$ because $Q_2$ has fixed frequency design and contribution by $Z$-field noise by it is negligible. The complete form of decoherence trace should be:
\begin{equation}
    \exp\left[-\frac{1}{2}(\Gamma_{1,Q1} + \Gamma_{1,Q2} + \Gamma_\Omega)\tau - \Gamma_{\varphi g}^2\tau^2\right].
\end{equation}
containing both relaxation and dephasing part of $g$-frame qubit. Below we discuss them separately.

\subsection{Relaxation of $g$-frame qubit}
\begin{figure}[bth]
    \centering
    \includegraphics[scale=0.7]{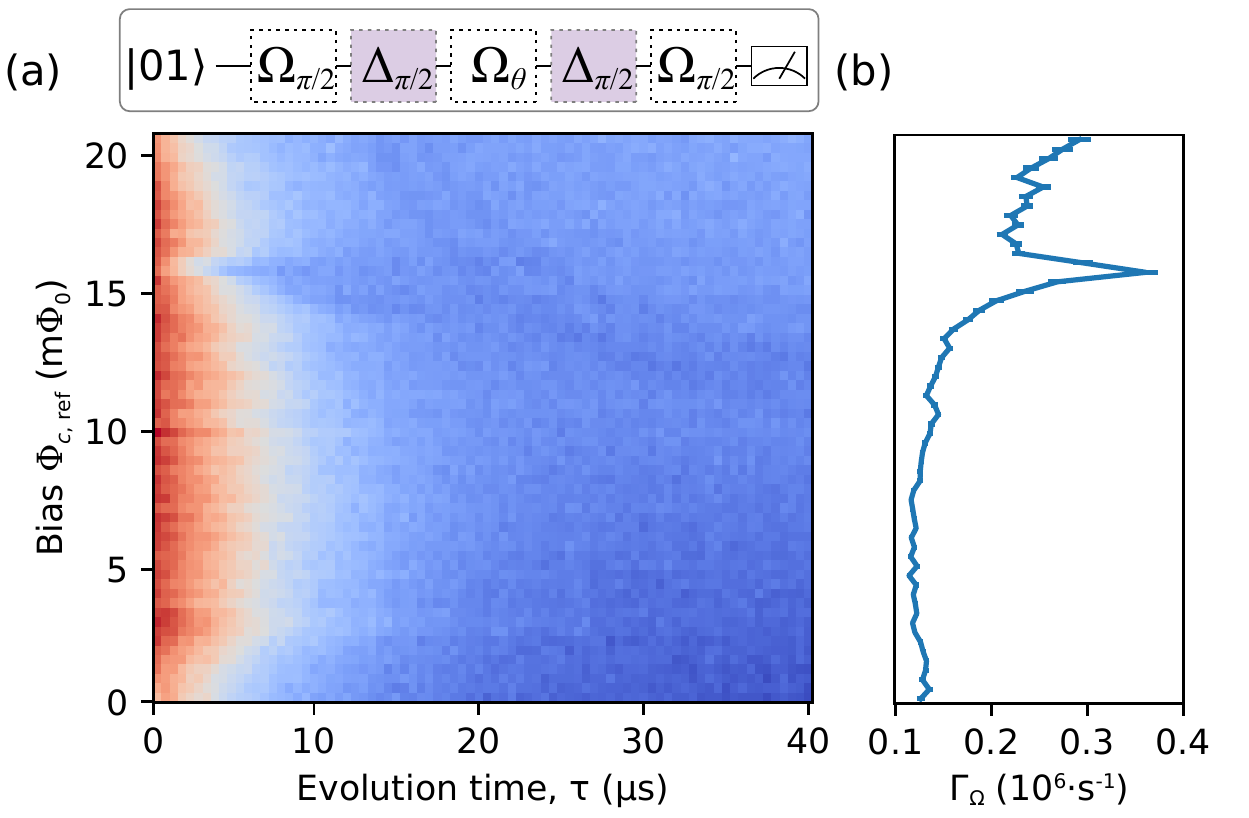}
    \caption{
        Relaxation of $g$-frame qubit at different bias. The operation sequence is shown on the top. (a). The decay traces (b). The fitted relaxation rate with errorbar of 1$\sigma$.
    }
\label{fig:t1g}
\end{figure}

Noise entering from tunable qubit detuning $\Delta$ acts as transverse noise. Fluctuation on frequency of either qubit is related to $\Delta$. With a non-zero Z-field, only those noise around $\Omega(\Phi_\mathrm{c})$ induce relaxation. Similar to single qubit, such noise leads a exponential relaxation according to Bloch-Redfield model, with a relaxation rate of $g$-frame qubit $\Gamma_\Omega\propto S_\Delta(\Omega)$. The relaxation of $g$-frame qubit is measured with experiment similar to single qubit case. The operation sequence and results are show in Figure.\ref{fig:t1g}.

\subsection{Dephasing of $g$-frame qubit}
\begin{figure}
    \centering
    \includegraphics[scale=0.7]{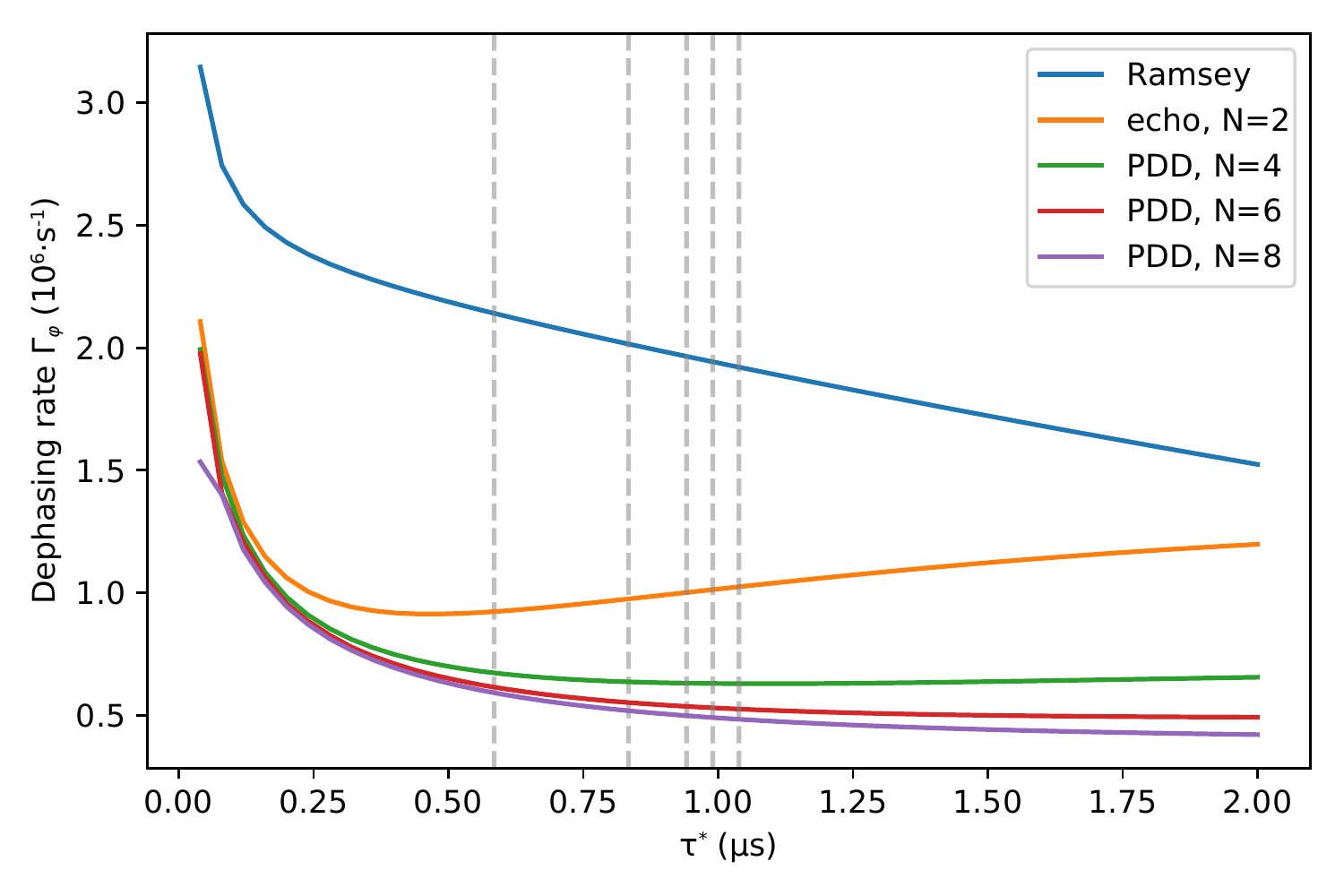}
    \caption{
        Calculated gaussian dephasing rate with different $\tau^*$. The dashed lines indicates the value of $\tau^*$ chosen for final calculation for different DD sequences.
    }
\label{fig:tau_star}
\end{figure}

\begin{figure}
    \centering
    \includegraphics[scale=0.7]{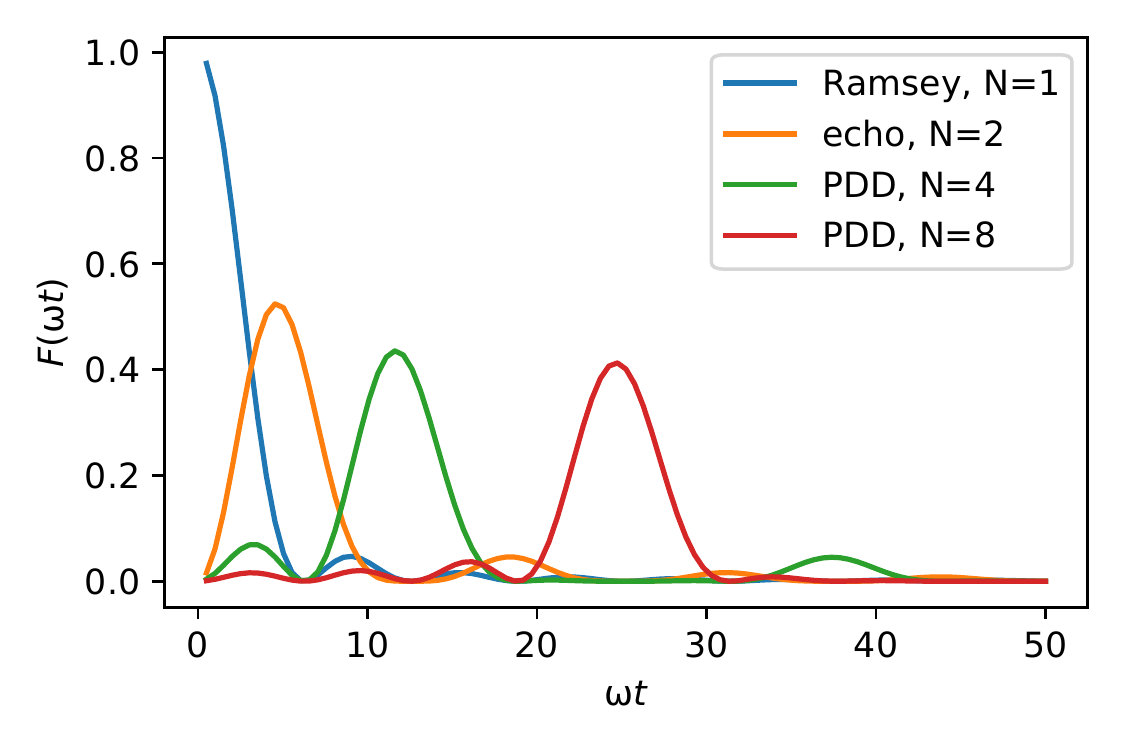}
    \caption{
        The filter function of DD sequence used in our experiment.
    }
\label{fig:filter_func}
\end{figure}

Dephasing is caused by low-frequency noise on the tunable coupling $\Omega(\Phi_\mathrm{c})$. In experiment, we turn on the coupling and tune qubits into resonance to expose the $g$-frame qubit to dephasing noise, during which the Hamiltonian is:
\begin{equation}
    \mathcal{H} = \frac{1}{2} \left( \Omega(\Phi_\mathrm{c}) + \delta\Omega(\Phi_\mathrm{c}) \right)\, \sigma_Z,
\end{equation}
and a relative phase is accumulated between $\frac{1}{\sqrt 2}(\ket{10}+\ket{01})$ and $\frac{1}{\sqrt 2}(\ket{10}-\ket{01})$ after time t:
\begin{equation}
    e^{-i\phi(t)} = e^{-i\int^t_0{[\Omega + \delta\Omega(t^\prime)] dt^\prime}} = e^{-i\Omega t}e^{-i\int^t_0{\delta\Omega(t^\prime) dt^\prime}}.
\end{equation}
Ensemble averaging the phase factor gives an real factor:
\begin{equation}
    \ev{e^{-i\phi(t)}} = e^{-\frac{1}{2}\ev{\phi^2(t)}} \equiv e^{-\chi(t)},
\end{equation}
above we assume the noise comes with statistics of Gaussian. $\chi(t)$ is so-called coherent integral, and can be written as:
\begin{equation}
    \chi(t) = \frac{1}{2}\ev{\left(\int_0^t{\delta\Omega(t^\prime)f(t;t^\prime)dt^\prime}\right)^2} = \int_0^\infty{\frac{\dd\omega}{2\pi}S_\Omega(\omega)\left|\tilde{f}(t;\omega)\right|^2},
    \label{eq:coherent integral1}
\end{equation}
where $S(\omega)$ is power spectrum density (PSD) of noise, the subscript donates noise source. In our experiment, fluctuation of coupling strength $\Omega$ is basically from flux noise. The PSD can be converted via sensitivity of system to noise by: $S_\Omega(\omega)=(\partial\Omega/\partial\Phi_\mathrm{c})^2S_{\Phi_\mathrm{c}}(\omega)\,$. $f(t;t^\prime)$ is a function with value of $+1$ or $-1$ for marking the polarity of accumulation at different time stage. $\tilde{f}$ is the Fourier transformation of it, following notation in \cite{cywinski_how_2008}. A pi pulse during the evolution filps the sign of $f$. Effect of DD sequences is specified by $f$. By defining a dimensionless function $F(\omega,t)\equiv|\tilde{f}|^2/t^2$ one can rewrite Eq.\ref{eq:coherent integral1} as:
\begin{equation}
    \chi(t)=t^2\left(\frac{\partial\Omega}{\partial\Phi_\mathrm{c}}\right)^2\int_0^\infty{\frac{\dd\omega}{2\pi}S_{\Phi_\mathrm{c}}(\omega)F(\omega,t)}.
    \label{eq:coherent integral2}
\end{equation}
$F(\omega,t)$ is called filter function. Its specific form of DD sequences used in our experiment is shown in Table.\ref{tab:filter function} and Fig.\ref{fig:filter_func}. Note that the $t$-dependence of filter function leads to non-gaussian decay trace, but for noise with $1/f$ like spectrum, the integral in actually insensitive to value of $t$ in $F(\omega,t)$ around the characteristic $1/e$ time (see Fig.\ref{fig:tau_star}). Thus one can replace it with a constant $\tau^*$ and the coherent integral is approximated as:

\begin{equation}
    \chi(t)=t^2\left(\frac{\partial\Omega}{\partial\Phi_\mathrm{c}}\right)^2\int_0^\infty{\frac{\dd\omega}{2\pi}S_{\Phi_\mathrm{c}}(\omega)F(\omega,\tau^*)}.
    \label{eq:coherent integral}
\end{equation}

We use this equation to compute the prediction line in maintext from a noise model. The $\tau^*$ is chosen as the fitted gaussian decay rate from experiment data.

\begin{table}[]
    \centering
    \begin{tabular}{ccc}
        \hline
        PDD & $F(\omega t)$ \\
        \hline
        N=1 (Ramsey) & $\frac{4}{\omega t^2}\sin^2{\frac{\omega t}{2}}$  \\
        N=2 (spin echo) & $\frac{16}{\omega t^2}\sin^4{\frac{\omega t}{4}}$  \\
        N=n (n is even) & $\frac{4}{\omega t^2}\tan^2{\frac{\omega t}{2n}}\sin^2{\frac{\omega t}{2}}$ \rule[-1.2ex]{0pt}{0pt}  \\
        \hline
    \end{tabular}
    \caption{Form of filter function of sequences used in our experiment. $n$ is number of intervals sliced by $\pi$ pulses.}
    \label{tab:filter function}
\end{table}

\subsection{Fitting decoherence traces}
\begin{figure}[bth]
    \centering
    \includegraphics[scale=0.7]{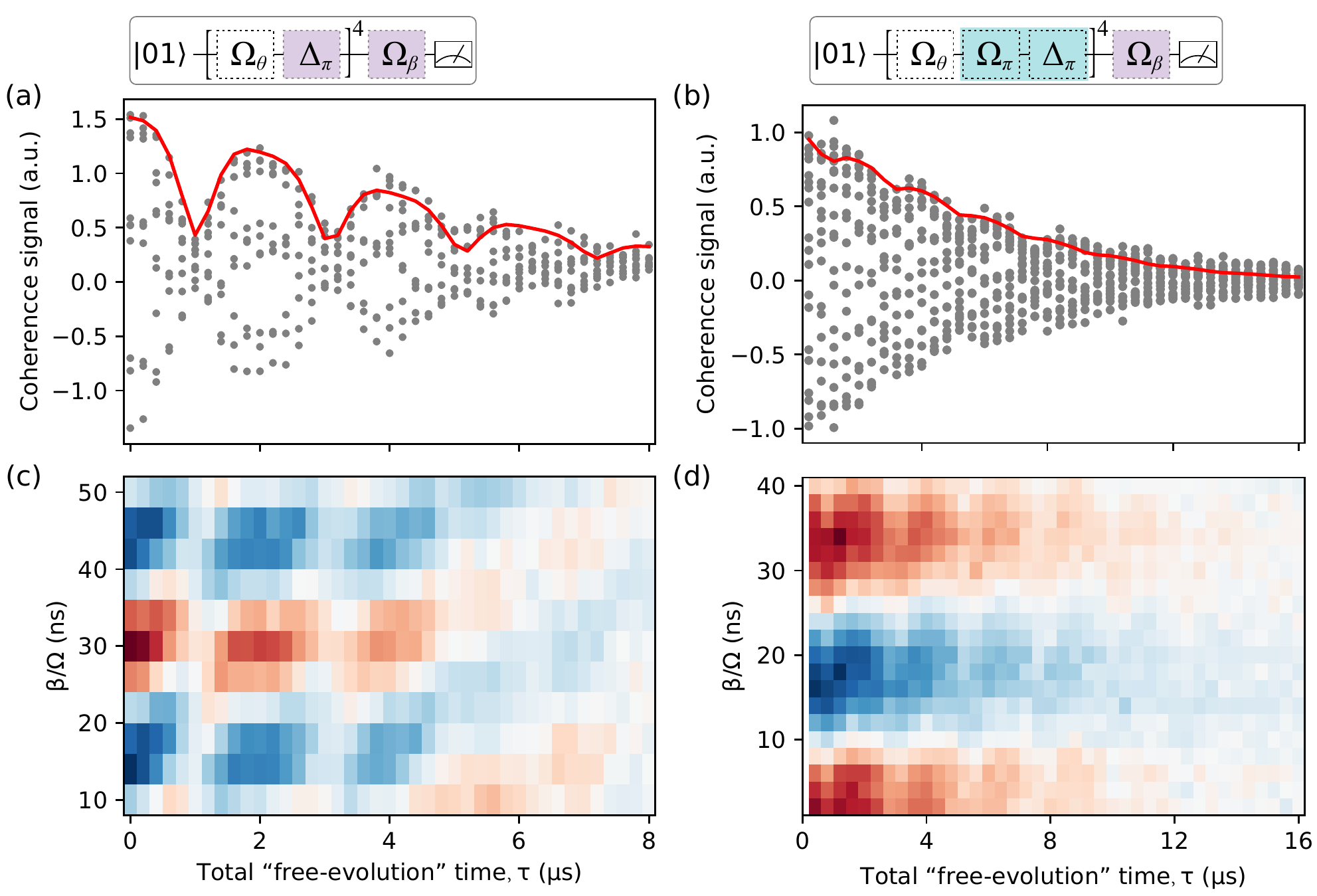}
    \caption{
        Measured population oscillation after the Ramsey-like experiment with DD sequence of (a), (c) X-4 and (b), (d) Y-4. The operation sequences are shown on the top. Beating feature found at LHS is basically removed in data shown at RHS. Note that Offset of oscillations are removed to compare against the fitted curve of dehpasing trace in (a), (b).
    }
\label{fig:fit_t2g}
\end{figure}

Decoherence of $g$-frame qubit is measured by a Ramsey-like experiment: First prepare state $\ket{01}$, which aligns to $\Delta$-axis. Then turn on coupling between qubits and let them swap population. Noise entering from $\Delta$-axis makes Bloch vectors spread out on surface of Bloch sphere, making the ensemble average of them lay inside the Bloch sphere. Measurement is performed by reading state of qubit 2, which gives the projection of Bloch vector on to $\Delta$-axis. In order to getting the length of Bloch vector, i.e. coherent amplitude, we perform a fast rotation along $\Delta$-axis on qubits, and record the oscillation of population in qubit 2. The data is shown as gray dots in \ref{fig:fit_t2g}. The oscillation amplitude is taken as a measure to coherent amplitude, and decreases with free evolution time. The decay trace is fitted with: $f(t)=A e^{-\Bar{\Gamma}_1 t-(\Gamma_{\varphi} t)^2}+B$ where $\bar{\Gamma}_1\approx0.18\times10^6\ \mathrm{s}^{-1}$ takes a fixed value of exponential decay rate among all fittings.

The decay trace shows some beatings at $\Omega<3\mathrm{MHz}$, which is attributed to low-frequency noise on qubits detune $\Delta$. Noise on $\Delta$-axis makes the eigen-axis of $g$-frame qubit tilts from $\Omega$-axis, evolving along the tilted axis making the projection of Bloch vector on $\Delta$-$Y$ plane encloses an ellipse, and hence leads to the beating pattern on the decay trace of measured oscillate amplitude. Such effect can also be removed with DD sequence. DD sequence refocuses out noise on perpendicular axis but has no effects on noise parallel to the axis where pi pulses is applied. So the beating feature shown with DD sequences with $\Delta_\pi$ pulses can be removed by those $Y_\pi$ pulse, as shown in Fig.\ref{fig:fit_t2g}.

\section{Noise from Room-Temperature Electronics}
\begin{figure}
    \centering
    \includegraphics[scale=0.7]{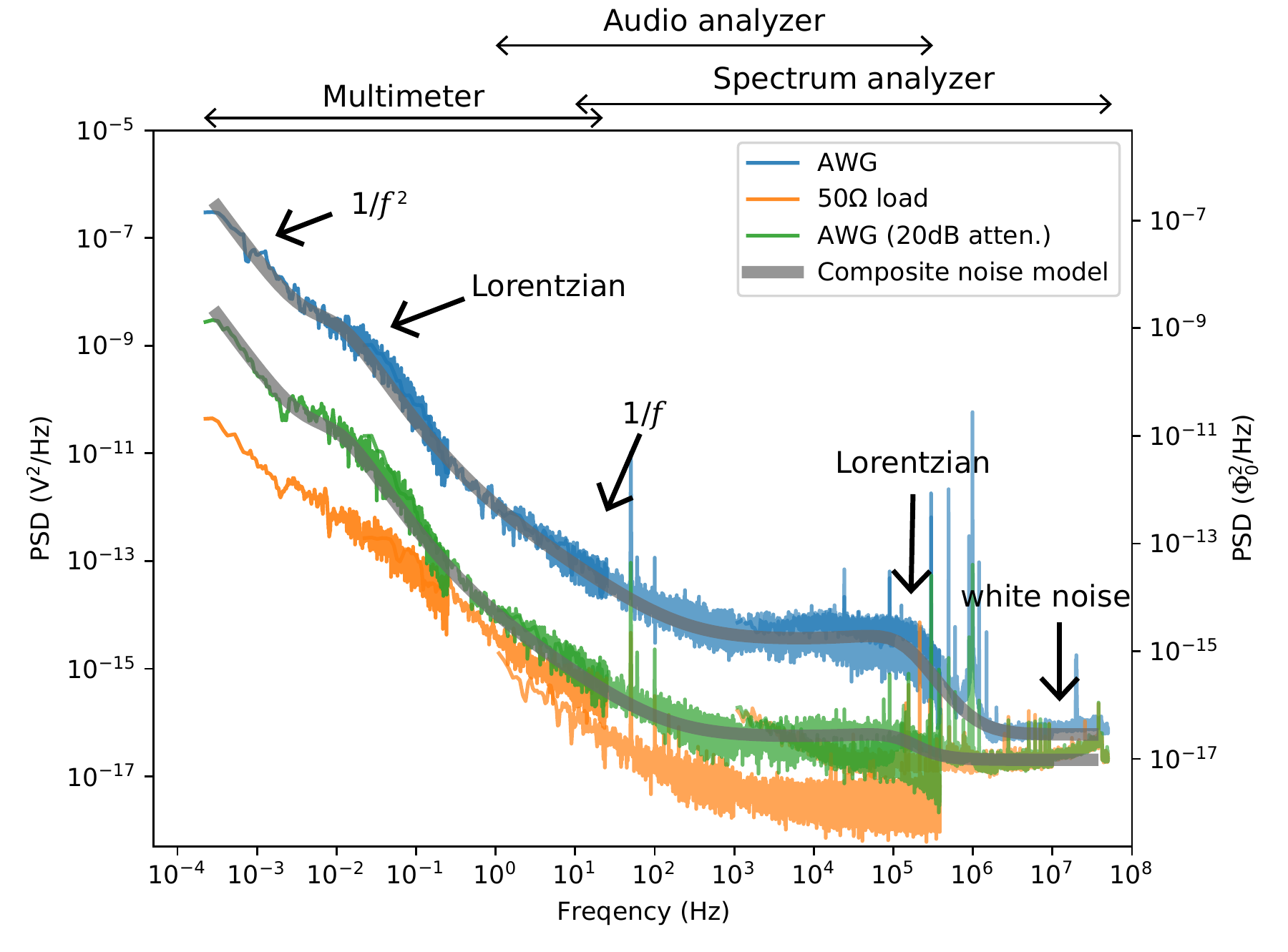}
    \caption{
        Noise spectrum at AWG output port used in experiment. The gray is a simplified model mimicking its behavior. The model is composited by five features capturing behavior of noise spectrum in different ranges. The voltage unit is converted into flux quanta unit with transfer function on bias line of coupler.
    }
\label{fig:noise_spec}
\end{figure}

In order to manipulating $g$-frame qubit, we use a arbitrary waveform generator (AWG) for generating pulses on coupler bias line to turn on and off the coupling. But noise on the AWG was brought to the device and cause dephasing of $g$-frame qubit in the mean time. Analysis above shows that this is a main source of noise. To quantitatively benchmark it, we measured noise signal on AWG output port, and computed its PSD. The result is shown in Fig.\ref{fig:noise_spec}.

The spectrum is measured by parts with several instruments: The part below 50Hz is measured by multimeter, part ranging from 1Hz to 300kHz is measured by an audio analyzer and part from 10Hz to 50MHz is measured by an spectrum analyzer. The later two instruments measure the PSD with a factor depending on the resolution bandwidth and the way analyzer works. The factor is excluded by scaling to be continuous with the multimeter result. The multimeter results are obtained by Fourier transform of a time series voltage value, each of them is averaged value over a integration time $dt$. According to Weiner-Khinchin theorem, the square of FFT results is related to to noise PSD with an factor $dt^2/T$. All the measurement output comes with an background by analyzer/multimeter itself. We measured the analyzer/multimeter background noise by connecting the probe to an $50\Omega$ load.

We use an simplified model to approximate the actual instrument noise PSD. The model is written: $A_1/\omega\ +\ A_2/\omega^2\ +\ A_\mathrm{l}\sigma_\mathrm{l}/((x-\mu_\mathrm{l})^2 - \sigma_\mathrm{l}^2)\ +\ A_\mathrm{h}\sigma_\mathrm{h}/((x-\mu_\mathrm{h})^2 - \sigma_\mathrm{h}^2) + A_\mathrm{w}$, the first two terms simulate the noise spectrum at low frequency and the two Lorentz peaks simulate the bulges at 150Hz and 50kHz, the last term is the background white noise. To achieve an estimation on Gaussian dephasing rate, we integrate $\int{\frac{\dd\omega}{2\pi}S(\omega)F(\omega\cdot \tau^*)}$ from 10mHz to 100MHz with $\tau^*$ given by dephasing time of system at sensitivity $\dd\Omega/\dd\Phi_\mathrm{c}=5\times10^9\, \mathrm{MHz}/\mathrm{m}\Phi_0$ from linear fits in maintext. We also take into account the noise by ground loop with form $A/\omega^{1.1}$ for the quick dephasing behavior at Y-1 sequence with attenuation. The results are shown as prediction lines in Fig.4 of maintext.

\bibliographystyle{naturemag}
\bibliography{references}